\documentclass[12pt]{iopart}
\usepackage{setstack}
\usepackage{graphicx}
\usepackage{amssymb}
\def\Painleve{Painlev\'e}
\def\Nordstrom{Nordstr\"om}
\begin{document}
%------------------------------------------------
\title{Heuristic approach to the Schwarzschild geometry}
%------------------------------------------------
\author{Matt Visser}
%------------------------------------------------
\address{School of Mathematical and Computing Sciences, 
Victoria University of Wellington, PO Box 600, Wellington, New Zealand}
%------------------------------------------------
\ead{matt.visser@mcs.vuw.ac.nz}
%------------------------------------------------
\begin{abstract}
%------------------------------------------------
  In this article I present a simple Newtonian heuristic for deriving
  a weak-field approximation for the spacetime geometry of a point
  particle. The heuristic is based on Newtonian gravity, the notion of
  local inertial frames [the Einstein equivalence principle], plus the
  use of Galilean coordinate transformations to connect the freely
  falling local inertial frames back to the ``fixed stars''. Because
  of the heuristic and quasi-Newtonian manner in which the spacetime
  geometry is obtained, we are at best justified in expecting it to be
  a weak-field approximation to the true spacetime geometry. However,
  in the case of a spherically symmetric point mass the result is
  coincidentally an \emph{exact} solution of the full vacuum Einstein
  field equations --- it is the Schwarzschild geometry in
  \Painleve--Gullstrand coordinates.
  
  This result is much stronger than the well-known result of Michell
  and Laplace whereby a Newtonian argument correctly estimates the
  value of the Schwarzschild radius --- using the heuristic presented
  in this article one obtains the entire Schwarzschild geometry. The
  heuristic also gives sensible results --- a Riemann flat geometry
  --- when applied to a constant gravitational field.  Furthermore, a
  subtle extension of the heuristic correctly reproduces the
  Reissner--{\Nordstrom} geometry and even the de~Sitter geometry.
  Unfortunately the heuristic construction is not truly generic. For
  instance, it is incapable of generating the Kerr geometry or
  anti-de~Sitter space.
  
  Despite this limitation, the heuristic does have useful pedagogical
  value in that it provides a simple and direct plausibility argument
  for the Schwarzschild geometry --- suitable for classroom use in
  situations where the full power and technical machinery of general
  relativity might be inappropriate. The extended heuristic provides
  more challenging problems --- suitable for use at the graduate
  level.

\vskip 0.50cm
\noindent
Dated: 23 September 2003; 11 March 2004; \LaTeX-ed \today
%------------------------------------------------
\end{abstract}
%------------------------------------------------
\pacs{gr-qc/0309072}
%------------------------------------------------
\maketitle
%------------------------------------------------
%Local defines
%------------------------------------------------
\def\d{{\mathrm{d}}}
\def\implies{\Rightarrow}
%------------------------------------------------
\def\eof{\Box}
%-------------------------------------------------------------------------
\newenvironment{warning}{{\noindent\bf Warning: }}{\hfill $\eof$\break}
\newenvironment{exercise}{{\noindent\bf Exercise: }}{\hfill $\eof$\break}
%-------------------------------------------------------------------------

%---------------------------------------------------
\section{Introduction}
%---------------------------------------------------

The heuristic construction presented in this article arose from
combining three quite different trains of thought:
\begin{itemize}
  
\item For an undergraduate course, I wanted to develop a reasonably
  clean motivation for looking at the Schwarzschild geometry suitable
  for students who had \emph{not} seen any formal differential
  geometry.  These students had however been exposed to Taylor and
  Wheeler's ``{\sl Spacetime Physics}''~\cite{Taylor1}, so they had
  seen a considerable amount of Special Relativity, including the
  Minkowski space invariant interval. They had also already been
  exposed to the notion of local inertial frames [local ``free-float''
  frames], which notion is equivalent to introduction of the Einstein
  equivalence principle. But there is no justification in the
  framework of~\cite{Taylor1} for introducing the Schwarzschild
  geometry.
  
\item Additionally, I was of course aware of the Newtonian idea of a
  ``dark star''; this notion going back to the Reverend John Michell
  (1783)~\cite{Michell,Lynden-Bell}, and popularized by Pierre Simon
  Marquis de Laplace (1799)~\cite{Laplace}.  Michell noted that in
  Newtonian physics the escape velocity from the surface of a star can
  exceed the speed of light when
\begin{equation}
  {1\over2} \; v_{\mathrm{escape}}^2 = {GM\over R} > {1\over2} \; c^2.
\end{equation}
That is, in Newtonian physics, (adopting the ``corpuscular
model''~\cite{Newton}), light cannot escape from the surface of a star
once
\begin{equation}
R < R_{\mathrm{escape}} = {2GM\over c^2},
\end{equation}
and this critical radius is (in suitable coordinates) \emph{exactly}
the same as the Schwarzschild radius of general relativity.

\item Finally, from exposure to the ``analogue models'' of general
  relativity~\cite{Unruh,Jacobson,Acoustic,Workshop}, I was aware of
  the large number of different ways in which effective Lorentzian
  spacetime geometries can arise in quite different physical systems.
  In particular, Bondi accretion~\cite{Bondi} (spherically symmetric
  accretion onto a gravitating point mass) leads to an ``acoustic
  geometry'' qualitatively similar to the Schwarzschild geometry.
\end{itemize}
By combining these ideas I found it was possible to develop a good
heuristic for the weak-field metric, which can be presented at a level
appropriate for undergraduate students. (Though some of the technical
comments made below are definitely not appropriate at this level.) The
remarkable feature of this heuristic is that for the Schwarzschild
geometry it happens to be \emph{exact}. This appears to be a
coincidence, but is a good way of introducing students who may not
intend to specialize in general relativity to the notion of a black
hole.

Now there is a long history of attempts using semi-Newtonian
plausibility arguments to ``derive'' approximate forms of the
Schwarzschild metric.  One of the earliest was that of Lenz (as
reported in~\cite{Jackiw} and~\cite{Sommerfeld}). Related but distinct
plausibility arguments have been developed by Schiff~\cite{Schiff} and
Harwit~\cite{Harwit}; though the physical basis of those attempts have
been questioned~\cite{Rindler,Gruber,Schutz2}.  Specifically, Rindler
develops a ``reductio ad absurdum'' argument by applying Schiff's
construction to a constant gravitational field~\cite{Rindler}, while
Gruber \emph{et al} develop an ``impossibility proof'' based on the
distinction between space curvature and gravitational
potential~\cite{Gruber}.

In contrast to the Lenz and Schiff plausibility arguments, the
heuristic developed in this article gives correct results for a
constant gravitational field, thus avoiding the criticism of Rindler,
and also side-steps the ``impossibility proof'' of Gruber \emph{et
  al}~\cite{Gruber}.  It does so by evading some of the basic choices
made in setting up those analyses. The key technical difference is the
use of off-diagonal components in the metric, and the fact that
``space'' (though not ``spacetime'') is exactly flat.  Though this is
``merely'' a coordinate change, it severely modifies the presentation,
analysis and conclusions of~\cite{Rindler,Gruber}.

%---------------------------------------------------
\section{The heuristic}
%---------------------------------------------------

%------------------------------------------------------------------
\subsection{Free float frames:}
%------------------------------------------------------------------

Start with a mass $M$ which has Newtonian gravitational
potential
\begin{equation}
\Phi = - {G M\over r}.
\end{equation}
Take a collection of local inertial frames [local free-float frames]
that are stationary out at infinity, and drop them.  In the Newtonian
approximation these local free-float frames pick up a speed
\begin{equation}
\vec v = - \sqrt{2GM\over r} \; \hat r.
\end{equation}
In the local free-float frames, physics looks simple, and the
invariant interval is simply given by the standard special relativity
result
\begin{equation}
\d s^2_{\mathrm{FF}} = - c^2 \; \d t_{\mathrm{FF}}^2 + \d x_{\mathrm{FF}}^2 + \d y_{\mathrm{FF}}^2 + \d z_{\mathrm{FF}}^2,
\end{equation}
where I want to emphasise that these are locally defined free-fall
coordinates.

%------------------------------------------------------------------
\subsection{Rigid frame:}
%------------------------------------------------------------------

Let us now try to relate these freely falling local inertial
coordinates to a rigidly defined surveyor's system of coordinates that
is tied down at spatial infinity --- that is, we want a coordinate
system connected to the ``fixed stars''.  Call these coordinates
$t_{\mathrm{rigid}}$, $x_{\mathrm{rigid}}$, $y_{\mathrm{rigid}}$, and $z_{\mathrm{rigid}}$.
Since we know the speed of the freely falling system with
respect to the rigid system, and we assume velocities are small, we
can write an approximate Galilean transformation:~\footnote{Note that
  the weak-field approximation implies the velocities of the local
  ``free fall'' frames is low, which is what permits us to use the
  Galilean approximation.}
\begin{equation}
\label{E:g1}
\d t_{\mathrm{FF}} = \d t_{\mathrm{rigid}};
\end{equation}
\begin{equation}
\label{E:g2}
\d\vec x_{\mathrm{FF}} = \d\vec x_{\mathrm{rigid}} - \vec v \; \d t_{\mathrm{rigid}}.
\end{equation}

\begin{warning}
  Most relativists will quite justifiably be concerned by the
  suggestion that there is a rigid background to refer things to.  The
  only reason we have for even hoping to get away with this is because
  all of the discussion is at this stage in the weak-field
  approximation.  For students with a special relativity background
  who have not been exposed to the mathematics of differential
  geometry the existence of these rigid coordinates is ``obvious'' and
  it is only the mathematically sophisticated students that have
  problems here.
\end{warning}

%------------------------------------------------------------------
\subsection{Approximate ``metric'':}
%------------------------------------------------------------------

Substituting, we find that in terms of the rigid coordinates the
spacetime interval takes the form
\begin{equation}
\label{E:g3}
\d s^2_{\mathrm{rigid}} = - c^2 \d t_{\mathrm{rigid}}^2 
+ \|\d\vec x_{\mathrm{rigid}} - \vec v \; \d t_{\mathrm{rigid}}\|^2.
\end{equation}
Expanding
\begin{equation}
\d s^2_{\mathrm{rigid}} = 
- [c^2 -v^2] \d t_{\mathrm{rigid}}^2  
- 2 \vec v \cdot \d \vec x \; \d t_{\mathrm{rigid}}
 +
\|\d\vec x_{\mathrm{rigid}}\|^2.
\end{equation}
That is
\begin{eqnarray}
\fl 
\d s^2_{\mathrm{rigid}} &=& 
- \left[c^2 -{2GM\over r}\right] \d t_{\mathrm{rigid}}^2  
+ 2 \sqrt{{2GM\over r}}\; \d r_{\mathrm{rigid}} \; \d t_{\mathrm{rigid}}
 +
\|\d\vec x_{\mathrm{rigid}}\|^2.
\end{eqnarray}
Now this is only an approximation --- we have used Newton's gravity,
Galileo's relativity, and the notion of local inertial frames. There
is no fundamental reason to believe this spacetime metric once $GM/r$
becomes large.

%------------------------------------------------------------------
\subsection{The miracle:}
%------------------------------------------------------------------

Dropping the subscript ``rigid'', the invariant interval
\begin{eqnarray}
\d s^2 &=& 
- \left[c^2 -{2GM\over r}\right] \d t^2  
+ 2 \sqrt{{2GM\over r}} \; \d r \; \d t
+\|d\vec x\|^2
\end{eqnarray}
is an \emph{exact} solution of {Einstein}'s equations of general
relativity, $R_{ab}=0$.  It is the Schwarzschild solution in disguise.
In spherical polar coordinates we have
\begin{eqnarray}
\fl
\d s^2 &=& 
- \left[c^2 -{2GM\over r}\right] \d t^2  
+ 2 \sqrt{{2GM\over r}} \; \d r \; \d t + \d r^2 
+r^2 \left[ \d\theta^2 +\sin^2\theta \d\phi^2 \right].
\end{eqnarray}
This is \emph{one} representation of the space-time geometry of a
Schwarzschild black hole, in a particular coordinate system (the
\Painleve--Gullstrand
coordinates)~\cite{Painleve,Gullstrand,Lemaitre}.  There are many
other coordinate systems you could use.

\begin{warning}
  I emphasise that this is a \emph{heuristic} that happens to give the
  exact result. I do not view this as a rigorous derivation of the
  Schwarzschild geometry from Newtonian physics, and on this issue
  disagree with reference~\cite{Czerniawski}. The heuristic does
  however provide a good motivation for being interested in this
  specific geometry, even if for pedagogical reasons you do not yet
  have the full vacuum Einstein equations available.
\end{warning}

\begin{exercise}
  More advanced students could at this stage be asked to find the
  coordinate transformation required to bring the above into standard
  curvature coordinate form (see for instance~\cite{Acoustic}):
\begin{equation}
t_{\mathrm{Schwarzschild}} = t - \left\{
2 r \; \sqrt{{2GM\over r\;c^2}} 
-
{4 G\,M\over c^3} \;\; \hbox{arctanh} \left(\sqrt{2G\,M\over r \; c^2}\right) 
\; \right\},
\end{equation}
or equivalently
\begin{equation}
\d t_{\mathrm{Schwarzschild}} 
= 
%\d t - 
%{\sqrt{2GM/(rc^2)}\over1-2GM/(rc^2)} \; {\d r\over c} 
%= 
\d t - 
{\sqrt{2GM/r}\over c^2-2GM/r} \; {\d r} 
.
\end{equation}
Furthermore, advanced students could also be asked to use a symbolic
algebra package to verify that the Ricci tensor is zero.
\end{exercise}

The \Painleve--Gullstrand coordinates are in fact equivalent to the
``rain frame'' introduced in Project B of the book ``Exploring black
holes''~\cite{Taylor2}. [See in particular equation (15) on page
B--13.] The underlying logic is rather different there however, as
those authors are presupposing that one has somehow found the exact
Schwarzschild solution, and are then trying to interpret it. In
contrast, the presentation of the current article could be used to
motivate interest in looking at the Schwarzschild solution.

%------------------------------------------------------------------
\subsection{Schwarzschild radius:}
%------------------------------------------------------------------

You can now easily see that something interesting happens at
\begin{equation}
{2GM\over r_S} = c^2;
\qquad\qquad
r_S = {2GM\over c^2};
\end{equation}
where we essentially recover the observations of Reverend John Michell
(1783) and Pierre Simon Marquis de Laplace (1799). In Einstein's
gravity the coefficient of $ \d t_{\mathrm{rigid}}^2 $ goes to zero at
the Schwarzschild radius; in Newton's gravity the escape velocity
\begin{equation}
v_{\mathrm{escape}} = \sqrt{2GM\over R}
\end{equation}
reaches the speed of light once $R=r_S$.

\begin{warning}
  This is a good point at which to introduce the students to the
  difference between ``coordinate velocity'' and ``physical
  velocity''. For ingoing and outgoing null rays we have
\begin{equation}
{\d r\over\d t} = - \sqrt{2GM\over r} \mp c.
\end{equation}
At the horizon the coordinate velocity of the infalling local inertial
frames (relative to the ``fixed'' coordinates) exceeds the speed of
light --- but this is perfectly acceptable in general relativity as it
is only a statement about coordinate systems, not a statement about
physical objects. The coordinate velocity of the outgoing light ray
goes to zero.  In addition, all physical velocities are limited by the
speed of light and must lie in or on the light cone defined by the
spacetime metric.
\end{warning}

\begin{warning}
  Some students will at this stage take the notion of a
  ``gravitational aether'' a little too seriously.  This is the major
  drawback of this heuristic, which can best be ameliorated by
  pointing out that this heuristic is not fundamental physics. The
  heuristic does not work well for the Reissner--{\Nordstrom} black
  hole, and needs a subtle patch.  More significantly, the heuristic
  fails utterly for the Kerr black hole.
\end{warning}

%---------------------------------------------------
\section{Constant gravitational field}
%---------------------------------------------------

Let us now consider a constant gravitational field in the $z$
direction, described by downward gravitational acceleration $g$. If we
now take a collection if inertial frames and drop them from any point
on the plane $z=0$ with initial velocity zero, then they will in the
Newtonian approximation pick up a speed
\begin{eqnarray}
\vec v =  -\sqrt{-2 \,g \,z} \; \hat z \qquad\qquad (z\leq 0).
\end{eqnarray}
Then switching from free fall coordinates [in which the metric is
simple] to ``rigid'' coordinates by using the Galilean transformation
of equations (\ref{E:g1})--(\ref{E:g2}) we find [using equation
(\ref{E:g3})]
\begin{eqnarray}
\fl 
\d s^2 &=& 
- \left[c^2 +2 \,g \,z\right] \d t_{\mathrm{rigid}}^2  
+ 2 \sqrt{{-2 \,g \,z}}\; \d z_{\mathrm{rigid}} \; \d t_{\mathrm{rigid}}
 +
\|\d\vec x_{\mathrm{rigid}}\|^2.
\end{eqnarray}
Again, I emphasise that this is only an approximation --- we have used
the Newtonian idea of a constant gravitational field, Galileo's
relativity, and the notion of local inertial frames. There is no
fundamental reason to believe this spacetime metric once $2\,g\,z/c^2$
becomes large.  Again, we encounter a ``miracle''. This metric is
again an exact solution of the full Einstein equations --- it is in
fact Riemann flat, as it must be in order to be compatible with the
full Einstein equivalence principle.

\begin{exercise}
  More advanced students could at this stage be asked to verify that
  this metric is Riemann flat. (That this metric is indeed Riemann
  flat may even surprise many researchers.)
\end{exercise}

Note that the metric above is in some sense the \Painleve--Gullstrand
equivalent of the Rindler wedge. Though everywhere Riemann flat, the
metric is real only for $z\leq 0$, and so it covers only part of the
Minkowski spacetime (similar to the situation for the Rindler wedge).

\begin{exercise}
  More advanced students could at this stage be asked to find the
  series of coordinate transformations required to bring the above
  into standard Rindler form. Initially, take
\begin{equation}
t_{\mathrm{Rindler}} = t_{\mathrm{rigid}} + \left\{
- {\sqrt{-2\,g\,z}\over g} +
 {c\over g} \;\; \hbox{arctanh} \left({\sqrt{-2g z}\over c}\right) 
\; \right\},
\end{equation}
or equivalently
\begin{equation}
\d t_{\mathrm{Rindler}} = 
\d t_{\mathrm{rigid}} - {\sqrt{-2\,g\,z}\over c^2+2\,g\,z} \; \d z,
\end{equation}
to obtain~\footnote{While not being useful for the purposes of the
  heuristic developed in the current article, it is perhaps
  interesting to note that Rindler's criticism of the Lenz and Schiff
  plausibility arguments~\cite{Rindler} is misleading in that the Lenz
  and Schiff construction applied to a constant gravitational field
  should lead to the metric (\ref{E:schiff-g}), not the standard
  Rindler wedge. It is only after several additional coordinate
  transformations that one recovers the Rindler wedge in the form
  (\ref{E:rindler}).}
\begin{equation}
\label{E:schiff-g}
\d s^2 = - [ c^2+2\,g\,z] \; \d t^2 + \d x^2 + \d y^2 
+ {c^2 \; \d z^2\over  c^2+2\,g\,z}
\end{equation}
Furthermore, advanced students could also be asked to use a symbolic
algebra package to verify that the Riemann tensor for this metric is
also zero.
\end{exercise}

\begin{exercise} 
  Show that in the metric (\ref{E:schiff-g}) the integral curves of
  the $t$ coordinate have position dependent 4-acceleration
\begin{equation}
a = {g \over 1 + 2\,g\,z/c^2}.
\end{equation}
In contrast the red-shifted ``local gravity''
\begin{equation}
\kappa = {\sqrt{|g_{tt}|}\over c} \; a = g
\end{equation}
is position independent. (Exactly the same phenomenon occurs in the
Rindler wedge.) It is in this sense that we are dealing with constant
gravitational acceleration.  The metric (\ref{E:schiff-g}) has a
horizon at $z=-c^2/(2g)$, where $a$ diverges, but $\kappa$ is well
behaved.  Indeed $\kappa_H = g$ is the surface gravity of that
horizon.
\end{exercise}

\begin{exercise} 
  To complete the transformation from (\ref{E:schiff-g}) to Rindler
  space, now perform two additional coordinate transformations
\begin{equation}
z \to z - {c^2\over 2g},
\end{equation}
to obtain
\begin{equation}
\d s^2 = - 2\,g\,z \; \d t^2 + \d x^2 + \d y^2 
+ {c^2 \;\d z^2\over  2\,g\,z},
\end{equation}
followed by
\begin{equation}
z \to {g\over2\; c^2}\;z^2,
\end{equation}
to obtain
\begin{equation}
\label{E:rindler}
\d s^2 = - {g^2\,z^2\over c^2}  \; \d t^2 + \d x^2 + \d y^2 
+ \d z^2.
\end{equation}
This finally is the usual representation of the Rindler wedge.
\end{exercise}

%---------------------------------------------------
\section{Discussion}
%---------------------------------------------------

Overall, I feel that the benefits of this heuristic outweigh the risks
--- once the specific spacetime geometry has been motivated in this
way, students not intending to specialize in general relativity can
simply be told that this is the Schwarzschild solution, and the
properties of this spacetime investigated in the usual
manner~\cite{Taylor2,Hartle,Schutz}.  Two key points are:
\begin{itemize}
  
\item This sort of argument cannot be fully general, even for weak
  fields.

\item That it is exact for Schwarzschild seems to be an accident. 
\footnote{%
  However, it should be noted that Laszlo Gergely~\cite{Gergely},
  adapting earlier work of Xanthopolous~\cite{Xanthopolous} has shown
  that there are certain situations in which a solution of the
  linearized weak-field Einstein equations can be bootstrapped into a
  solution of the full nonlinear Einstein equations. In the current
  heuristic the details are different, but the flavour of the result
  seems similar.}

\end{itemize}  
I expand on these points below. Some of the issues raised below are
very definitely nontrivial and not suitable for an undergraduate
audience. Suitably modified, some points may be of interest for
mathematically sophisticated students who do not have a significant
physics background.

%---------------------------------------------------
\subsection{Spherical symmetry:}
%---------------------------------------------------
That the heuristic presented above, or some variant thereof, has some
chance of working for general time independent [stationary]
spherically symmetric geometries, can be seen by appropriately
choosing the coordinates. Stationarity plus spherical symmetry is
enough to yield~\cite{Schutz}
\begin{equation}
\d s^2 = - A(r) \;\d t^2 + 2 B(r) \; \d r \, \d t + E(r) \; \d r^2 
+ F(r,t) \; \d \Omega^2.
\end{equation}
The usual procedure at this point is to use the coordinate freedom in
the $r$-$t$ plane to eliminate the off-diagonal term, and also to
normalize the $\d\Omega^2$ coefficient, to locally obtain the
manifestly static form
\begin{equation}
\label{E:diag}
\d s^2 = - A(r)\;  \d t^2 + E(r) \; \d r^2 + r^2\; \d \Omega^2.
\end{equation}

\begin{warning}
  Coordinate arguments will only tell you that you can do this in
  suitably defined local coordinate patches. That global coordinate
  systems of this type exist for stars is a deep result that requires
  some assumptions about the the regularity of the centre, the nature
  of matter, and dynamical information from the Einstein equations.
  Specifically, if the null energy condition holds then there are no
  ``wormhole throats'' and the coordinate $r$ is continuously
  increasing as one moves away from the
  center~\cite{Morris-Thorne,Lorentzian}.
\end{warning}

In contrast, starting from (\ref{E:diag}) one could define a new time
coordinate by
\begin{equation}
\d t_{\mathrm{new}} = \d t_{\mathrm{old}} \pm \sqrt{{E(r)-1\over A(r)}} \; \d r,
\end{equation}
and so obtain~\cite{Lake,Martel,Husain,Kraus}
\begin{equation}
\label{E:pg}
\d s^2 = - A(r) \;\d t^2 \pm 2 \sqrt{{E(r)-1\over A(r)}} \; \d r \, \d t + \d r^2 
+ r^2 \; \d \Omega^2.
\end{equation}
There is a technical restriction here, that the $E(r,t)$ occurring in
(\ref{E:diag}) above be greater than unity. Otherwise one will
encounter imaginary metric components in (\ref{E:pg}). This issue is
relevant deep inside a Reissner--\Nordstrom{} black hole, or more
prosaically, in anti-de Sitter space.

One then defines functions $N(r,t)$ and $\beta(r,t)$ so that
\begin{equation}
\d s^2 = 
- [N(r,t)^2 - \beta(r,t)^2]  \;\d t^2 - 2  \beta(r,t) \; \d r \, \d t 
+ \d r^2  + r^2 \; \d \Omega^2,
\end{equation}
implying
\begin{equation}
\d s^2 = 
- N(r,t)^2\;\d t^2 
+ (\d r -\beta(r,t)\;\d t)^2 + r^2 \; \d \Omega^2,
\end{equation}
The interpretation is that in spherical symmetry one can almost always
[patch-wise] choose coordinates to make the spatial slices [though
\emph{not spacetime}] flat. In the language of the ADM decomposition
(see for instance~\cite{Lorentzian,MTW}), you bury all of the
spacetime curvature in the lapse and shift functions, $N(r,t)$ and
$\beta(r,t)$.  The heuristic presented above consists of setting
$N(r,t)=c^2$ and $\beta(r,t) = - \sqrt{2GM/r}$, but we now see that by
instead choosing suitable ansatze for the lapse and shift we would be
able to fit a wide class of spherically symmetric spacetimes.

Coordinates of this type are known as \Painleve--Gullstrand
coordinates~\cite{Painleve,Gullstrand,Lemaitre} and have many
pedagogically and computationally useful
properties~\cite{Lake,Martel,Husain,Kraus}.  A particularly nice
feature is that infalling observers cross the horizon in finite
coordinate time, so that one does not have to confront the
pseudo-paradox encountered in standard coordinates where one has to
wait an infinite amount of coordinate time (but finite proper time) in
order for a test particle to reach the horizon.

Historically the \Painleve--Gullstrand coordinates were developed in
an attempt to show there was something wrong with the Schwarzschild
coordinates~\cite{Painleve,Gullstrand}. (More recently, see
also~\cite{Czerniawski}.) However, as emphasised by
Lema\^\i{}tre~\cite{Lemaitre}, these are just a specific choice of
coordinates [albeit somewhat unusual ones] and their adoption or
rejection cannot affect the underlying physics or mathematics.

\medskip
The heuristic applied to a generic spherically symmetric field yields
\begin{eqnarray}
\fl
\d s^2 &=& 
- \left[c^2 + 2\Phi(r)\right] \d t^2  
+ 2 \sqrt{-2\Phi(r)} \; \d r \; \d t + \d r^2 
+r^2 \d \Omega^2.
\end{eqnarray}
If there is a well defined surface beyond which the object is vacuum,
then in that region Newtonian physics gives $\Phi(r)=-GM/r$ and so our
heuristic reproduces the Birkhoff theorem~\cite{MTW}.  But in general,
in Newtonian gravity the gravitational acceleration in a situation
with spherical symmetry is
\begin{equation}
\vec g = - {G \,m(r)\over r^2} \; \hat r.
\end{equation}
Integrating, this now implies
\begin{equation}
\Phi(r) = \int g \; \d r = - { G \,m(r)\over r} + G \int \rho(r)\; r \; \d r
\end{equation}
As long as the density falls off sufficiently rapidly at spatial
infinity, $\rho(r)\to C/r^{3+\epsilon}$, the second term is
sub-dominant near spatial infinity, $\int\rho(r)\;r\;\d r \to C/
r^{1+\epsilon}$, and we can (in the weak field limit) write
\begin{eqnarray}
\fl
\d s^2 &=& 
- \left[c^2 -{2G\,m(r)\over r}\right] \d t^2  
+ 2 \sqrt{{2G\,m(r)\over r}} \; \d r \; \d t + \d r^2 
+r^2 \left[ \d\theta^2 +\sin^2\theta \d\phi^2 \right].
\end{eqnarray}
This geometry, while reasonably general, is \emph{not} the most
general weak-field metric possible in general relativity. For this
reason our heuristic will \emph{not} be able to exactly reproduce
\emph{all} spherically symmetric geometries. [You could also come to a
similar conclusion, but without some of the interesting intermediate
results, by noting that the general spherically symmetric geometry is
specified by two arbitrary functions $N(r,t)$ and $\beta(r,t)$ whereas
the heuristic depends on only one arbitrary function $\Phi(r,t)$.]

%---------------------------------------------------
\subsection{Reissner--{\Nordstrom} geometry:}
%---------------------------------------------------

The exact Reissner--{\Nordstrom} geometry~\cite{MTW} corresponds to
the choice $N(r,t)=c^2$ and 
\begin{equation}
\beta(r,t) = - \sqrt{2GM/r-Q^2/r^2}
\end{equation}
so that
\begin{eqnarray}
\label{E:RN}
\fl
\d s^2 &=& 
- \left[c^2 -{2GM\over r}+{Q^2\over r^2}\right] \d t^2  
+ 2 \sqrt{{2GM\over r}-{Q^2\over r^2}} \; \d r \; \d t + \d r^2 
+r^2 \d \Omega^2.
\end{eqnarray}
Unfortunately, while we can put the Reissner--{\Nordstrom} geometry
into the \Painleve--Gullstrand \emph{form} appropriate for our
heuristic analysis, the precise details do not quite mesh with the
most naive form of the heuristic.  For a charged particle surrounded
by an electric field we could argue that the equivalence of mass and
energy requires
\begin{equation}
\rho 
=  M \; \delta^3(\vec x) +  {1\over8\pi} E^2
=  M \; \delta^3(\vec x) + {1\over8\pi} {Q^2\over r^4},
\end{equation}
so that
\begin{equation}
m(r) = M - {1\over 2} {Q^2\over r}.
\end{equation}
Unfortunately this now implies
\begin{equation}
\fl
\Phi 
= -\int_r^\infty g(\tilde r) \; \d \tilde r 
= - G \int_r^\infty \left[ {M\over \tilde r^2} 
-  {1\over 2} {Q^2\over \tilde r^3} \right] \; \d \tilde r
= - G \left[{M\over r} - {Q^2\over 4r^2} \right],
\end{equation}
and the coefficient of the $Q^2$ term does not match the exact
Reissner--{\Nordstrom} geometry, with a missing factor of $2$.  Though
the naive heuristic does not exactly reproduce the
Reissner--{\Nordstrom} geometry, it does get remarkably close.

There is a slightly less naive version of the heuristic that does the
job: If we linearize the full Einstein equations around flat space
then for a static situation the linearized equations imply
\begin{equation}
\nabla^2 \Phi = G \left\{ \rho + \sum_i {p_i\over c^2} \right\},
\end{equation}
where the $p_i$ are the principal pressures. (Of course, deriving this
result properly requires exactly the sort of technical analysis that I
had hoped to avoid by adopting the heuristic. For non-technical
students, one could simply assert that pressures contribute to the
gravitational field in the same way that density does.) Spherical
symmetry, plus the tracelessness of the electromagnetic stress-energy
tensor now implies that in the radial and transverse directions
\begin{equation}
\rho = - {p_r\over c^2} = {p_t\over c^2},
\end{equation}
so that (away from the central delta function)
\begin{equation}
\rho + \sum_i {p_i\over c^2} \to 2 \rho.
\end{equation}
This factor of $2$ now compensates the ``missing'' factor of $2$
above, and with this extension to the heuristic we exactly reproduce
the Reissner--{\Nordstrom} geometry.

\begin{warning}
  This extension of the heuristic brings in ideas that are
  considerably more subtle and advanced than those needed for the
  simple Schwarzschild solution. (At a minimum, you would need to
  motivate the idea that pressures and tensions also generate
  gravitational fields, and would need to be able to rely upon the
  student's prior exposure to the Maxwell electromagnetic stress
  tensor.)
\end{warning}

\begin{warning}
  In addition the Reissner--{\Nordstrom} geometry in
  \Painleve--Gullstrand coordinates suffers from the unpleasant
  feature that the shift vector becomes imaginary for $r<Q^2/2GM$.
  Fortunately this occurs inside the inner horizon [the Cauchy
  horizon] where we should not be trusting the geometry in any case
  [because the Cauchy horizon is unstable to any infalling
  stress-energy]. This type of coordinate singularity can be avoided
  by generalizing the form of the metric, see for
  instance~\cite{Volovik}, but this moves us outside the framework of
  the Newtonian heuristic, and is another reason for viewing the
  Newtonian heuristic as non-fundamental.
\end{warning}

%---------------------------------------------------
\subsection{de Sitter geometry:}
%---------------------------------------------------

Once you have adopted the extended heuristic appropriate for the
Reissner--{\Nordstrom} geometry, a nice feature is that it
automatically works for positive cosmological constant as well.  The
stress-energy equivalent to any cosmological constant satisfies
\begin{equation}
\rho = - {p_i\over c^2},
\end{equation}
so that
\begin{equation}
\rho + \sum_i {p_i\over c^2} \to  -2 \rho.
\end{equation}
This factor of $-2$ now implies (if $\rho$ is positive) a repulsive
force, away from the origin. A local free float frame, \emph{initially
  dropped at the origin} with velocity zero, will accelerate outwards at a rate
\begin{equation}
g = {2 G m(r)\over r^2} = {8\pi\over3} \rho r,
\end{equation}
and pick up a speed
\begin{eqnarray}
{1\over2} v^2 ={4\pi\over3} G \; \rho \; r^2,
\end{eqnarray}
whence we obtain
\begin{eqnarray}
\label{E:de-sitter}
\d s^2 &=& 
- \left[c^2 -  {8\pi\over3} G\, \rho \; r^2\right] \d t^2  
+ 2 \sqrt{ {8\pi\over3} \; G\,\rho  } \; r\; \d r \; \d t + \d r^2 
+r^2 \d \Omega^2.
\end{eqnarray}
Though the extended heuristic has now reproduced the de Sitter
solution, note that the free float frames are now dropped form the
origin --- not spatial infinity. This is a symptom of the fact that
the heuristic is not well adapted to dealing with geometries that are
not asymptotically flat.

\begin{warning}
If one now attempts to apply the heuristic to the Schwarzschild--de
Sitter [Kottler] geometry, one has to face the very basic question of
where the free float frames should be dropped from. There really is no
good answer to this. Worse, if one attempts to deal with anti-de Sitter
space, then the speed of the free float frames is everywhere
imaginary --- again the heuristic breaks down and we have another
reason for viewing the Newtonian heuristic as non-fundamental.
\end{warning}

%---------------------------------------------------
\subsection{Kerr geometry:}
%---------------------------------------------------

The heuristic approach definitely fails for the Kerr geometry --- most
fundamentally because the Kerr geometry is not spherically symmetric.
More technically, the \Painleve--Gullstrand coordinates require the
existence of flat spatial slices, and the Kerr geometry does not
possess such a slicing. In fact the Kerr geometry does not even
possess a conformally flat spatial slicing~\cite{Garat,Kroon,Kroon2}.
The closest that one seems to be able to get to \Painleve--Gullstrand
coordinates seems to be Doran's form of the metric~\cite{Doran}, for
which a brief computation shows that $N(r,\theta)=c^2$; the lapse
function is a \emph{constant} independent of position. Unfortunately
the spatial slices in Doran's coordinates are very definitely not
flat. More critically I have not been able to find \emph{any} useful
set of coordinates that would make the Kerr geometry amenable to
treatment along the lines of the heuristic approach considered above.
For this reason, among others, the heuristic approach should not be
thought of as fundamental physics.

%---------------------------------------------------
\subsection{Bondi acoustic geometry:}
%---------------------------------------------------

In contrast, a particularly nice feature of the heuristic analysis is
the clean relationship with the acoustic geometry occurring in Bondi
accretion~\cite{Bondi}.  Consider a fluid with a linear equation of
state
\begin{equation}
\rho(p) = \rho_0 + {p \over c_s^2}; \qquad   p = (\rho-\rho_0) \; c_s^2;
\end{equation}
undergoing spherically symmetric accretion onto a compact
object~\cite{Bondi}. Here $c_s$ is the speed of sound, assumed
constant. Then as long as back-pressure can be neglected, the infalling
matter satisfies $v = - \sqrt{2GM/r} \; \hat r$.  Sound waves travelling
on the background of this infalling matter will then travel at speed
\begin{equation}
\left\|  - \sqrt{2GM/r} \; \hat r + c_s \; \hat n \right\|
\end{equation}
with respect to the fixed stars. This situation is tailor-made for
application of the acoustic geometry
formalism~\cite{Acoustic,Workshop}, and as long as the back-pressure is
negligible the effective acoustic geometry is exactly the
Schwarzschild geometry with the speed of light replaced by the speed
of sound; that is, with the substitution $c\to c_s$.

%---------------------------------------------------
\subsection{Spatially flat geometries:}
%---------------------------------------------------

Recently Nurowski, Sch\"ucking, and Trautman have used metrics with
flat spatial slices (which include as a subset the spherically
symmetric geometries in \Painleve--Gullstrand coordinates) to
investigate general relativistic spacetimes with close Newtonian
analogues~\cite{Trautman}.  That approach, since it starts from the
full Einstein equations, is in some sense the converse of the
heuristic developed here. Metrics with flat spatial slices also occur
ubiquitously in the various ``analogue model'' geometries, not just
the spherically symmetric ones. A necessarily incomplete set of
references
includes~\cite{Unruh,Jacobson,Acoustic,Workshop,Analog,Uwe,Fedichev}.
The class of spatially flat geometries appears to be of interest in
its own right, even if it is not general enough to contain the Kerr
geometry.

%---------------------------------------------------
\subsection{Summary:}
%---------------------------------------------------

The basic heuristic discussed in the first few pages of this article
can easily be explained to undergraduate students who have no
intention of specializing in general relativity, and can be used to
motivate interest the Schwarzschild geometry and black hole physics.
The remarkable feature of the heuristic is that it leads directly to
what is certainly the physically most important exact solution of the
full Einstein equations --- the Schwarzschild geometry (albeit in
\Painleve--Gullstrand coordinates).  As we have seen in the
commentary, this leads naturally to a number of rather more technical
issues and questions (suitable for graduate student problems) hiding
in this rather innocent looking heuristic.  Though the heuristic
should in no way be thought of as fundamental physics, it does have
considerable pedagogical value.

%---------------------------------------------------
\appendix
%---------------------------------------------------

%------------------------------------------------
\ack
%------------------------------------------------

This Research was supported by the Marsden Fund administered by the
Royal Society of New Zealand. I wish to thank Jan Czerniawski, Laszlo
Gergely, Roman Jackiw, Edwin Taylor, and Grigori Volovik for their
comments and questions.

%------------------------------------------------
\section*{References}
%------------------------------------------------

%------------------------------------------------

%--------------------------------------------------- 

\begin{thebibliography}{99}
\bibitem{Newton} Isaac Newton, {\sl Optics}, 1704:
\begin{quote}
  {\sl Query 1:} And do not Bodies act upon Light at a distance and,
  by their action, bend its Rays, and is not this action strongest at
  the least distance?
\end{quote}
\end{thebibliography}
\end{document}